\documentclass[aps,nofootinbib]{revtex4}%
\usepackage{amsfonts}
\usepackage{amsmath}
\usepackage{amssymb}
\usepackage{graphicx}%
\providecommand{\U}[1]{\protect\rule{.1in}{.1in}}

\begin{document}
\title{Traversing Cosmological Singularities\\Complete Journeys Through Spacetime Including Antigravity \footnote{Invited
article, to appear in \textit{Beyond the Big Bang, Competing Scenarios for an
Eternal Universe}, Ed. Rudy Vaas, ISBN 978-3-540-71422-4,
http://www.springer.com/astronomy/general+relativity/book/978-3-540-71422-4 .
}}
\author{Itzhak Bars}
\affiliation{Department of Physics and Astronomy, University of Southern California, Los
Angeles, CA 90089-0484, USA}

\begin{abstract}
A unique description of the Big Crunch-Big Bang transition is given at the
classical gravity level, along with a complete set of homogeneous, isotropic,
analytic solutions in scalar-tensor cosmology, with radiation and curvature.
All solutions repeat cyclically; they have been obtained by using conformal
gauge symmetry (Weyl symmetry) as a powerful tool in cosmology, and more
generally in gravity. The significance of the Big Crunch-Big Bang transition
is that it provides a model independent analytic resolution of the
singularity, as an unambiguous and unavoidable solution of the equations at
the classical gravitational physics level. It is controlled only by geometry
(including anisotropy) and only very general features of matter coupled to
gravity, such as kinetic energy of a scalar field, and radiation due to all
forms of relativistic matter. This analytic resolution of the singularity is
due to an attractor mechanism created by the leading terms in the cosmological
equation. It is unique, and it is unavoidable in classical relativity in a
geodesically complete geometry. Its counterpart in quantum gravity remains to
be understood.

\end{abstract}
\maketitle


\section{Introduction and Summary of Results}

Resolving the big bang singularity is one of the central challenges for
fundamental physics and cosmology. The issue arises naturally in the context
of classical relativity due to the geodesically incomplete nature of the
singular spacetime \cite{hawking1}-\cite{hawkPenrose}. The resolution may not
be completed until the physics is understood in the context of quantum
gravity. String theory or other forms of quantum gravity have not yet reached
the technical level to tackle this cosmological issue. Solutions in the
classical theory of relativity coupled to a scalar field are directly
connected to the quantum theory in the context of string theory, where the
classical field theory solution provides a critical string background
consistent with perturbative worldsheet conformal symmetry. Therefore analytic
solutions obtained in the classical theory of relativity, and insights on
geodesic completeness with the resolution of the singularity, are relevant not
only for the classical theory but also contribute to the quantum treatment.
Indeed such classical solutions may well guide the future development of the
quantum theory of gravity as applied to cosmology.

In recent papers with Shih-Hung Chen, Paul Steinhardt and Neil Turok
\cite{inflation-BC}-\cite{BCST2}, I studied a model of a scalar field
minimally coupled to gravity, with some specific forms of potential energy for
the scalar field, and included curvature and radiation as two additional
parameters. Our goal was to obtain analytically the complete set of
configurations of a homogeneous and isotropic universe as a function of time.
This led to a geodesically complete description of the universe, including the
passage through the cosmological singularities, at the classical level. We
gave all the solutions analytically without any restrictions on the parameter
space of the model or initial values of the fields.

The overall approach in this investigation was in contrast to specific
analytic, approximate, or numerical solutions that are usually fine tuned from
the point of view of initial conditions and/or the potential energy function
$V\left(  \sigma\right)  $~for the scalar field, to force a solution in which
the universe has a particular desired behavior as motivated by prejudices and
observations. Instead, we aimed at the global structure of solution space that
can emerge from a class of theories, so that we can gain a better
understanding of how the features of our own universe could emerge. Indeed we
learned important general lessons that transcend the specific forms of
potentials $V\left(  \sigma\right)  $ we could analyze analytically.

We found that geodesic completeness together with anisotropy of spacetime play
very significant roles in understanding the phenomena in the vicinity of the
Big Bang or Big Crunch singularities, and the transition between them, which
involves a brief period of antigravity. Therefore it is important to describe
the entire set of geodesically complete analytic solutions first in the
absence of anisotropy and then their modification in the presence of anisotropy.

In the \textit{absence} of anisotropy, we could solve the equations fully only
for some forms of the potential energy of the scalar field, but this was
sufficient to extract general lessons that would apply to a much wider class
of models. An essential ingredient was the requirement of a geodesically
complete geometry. The complete set of exact solutions, combined with
conformal symmetry, provided the ingredients to construct the geodesically
complete geometry. Then we learned that for generic solutions (meaning no
restrictions on initial conditions or parameters of the model), the universe
goes through a singular (zero-size) bounce by entering a period of antigravity
at each big crunch and exiting from it at the following big bang. This happens
cyclically without violating the null energy condition (Fig.2a). There is a
special subset of geodesically complete non-generic solutions which perform
zero-size bounces without entering the antigravity regime in all cycles. For
these, initial values of the fields are synchronized and quantized but the
parameters of the model are not restricted (Figs.4-7). There is also a subset
of spatial curvature-induced solutions that have finite-size bounces in the
gravity regime and never enter the antigravity phase (Figs.8,9). These exist
only within a small continuous domain of parameter space without fine tuning
initial conditions. To obtain these results, we identified 25 regions of a
6-parameter space in which the complete set of analytic solutions were
explicitly obtained.

In the \textit{presence} of anisotropy, all solutions mentioned
above are strongly modified near the singularity (see Figs.2b,3). An
attractor mechanism pulls the solution to a particular region of the
phase space of the degrees of freedom (scalar and tensor fields).
The dominant terms in the cosmological equations are the anisotropy
of the geometry, and the kinetic energy of the scalar field, while
\textquotedblleft radiation\textquotedblright\ due to all species of
relativistic particles is the next most important ingredient. Other
terms, such as inhomogeneities, massive matter, potential energy,
cosmological constant, etc. are all subdominant near the
singularity. In particular, since the potential energy of the scalar
field is negligible (unless it has non-typical behavior near the
singularity), the resolution of the singularity described here is
essentially model independent. With only the dominant terms just
mentioned, we discovered a new phenomenon which is unambiguous
classical evolution through cosmological singularities, as well as
generic in geodesically complete spacetimes. This is the attractor
mechanism that modifies all the solutions described in the previous
paragraph, irrespective of their initial conditions, such that it
ensures the generic and special solutions to undergo a big
crunch/big bang transition by contracting to zero size, passing only
through the \textit{origin of field space} (see figures) to enter a
brief antigravity phase, shrinking to zero size again, and
re-emerging into an expanding normal gravity phase, again by passing
only through the origin of field space, as in Figs.2b,3. The origin
of field space is then dynamically \textit{forced to be a natural
initial condition} for all scalar fields at the Big Bang.

This result of the attractor can be used as an unavoidable initial condition
for all scalar fields, including the dilaton of string theory, the Higgs
field, or other scalar fields, in order to discuss their cosmological history.
In fact, based on this observation, the history of the Higgs field is
currently under construction \cite{BCST3}. Similarly, this may be useful for
the construction of complete cosmologies like the inflation model
\cite{inflationGuth}-\cite{SciAm}, the cyclic model \cite{cyclic1}%
-\cite{Rendall}, or a hybrid model that combines the attractive features of
both scenarios.

New tools to obtain analytic solutions of the Friedmann equations used in our
work came from Two-Time Physics (2T-physics) that was developed since 1995
(for a recent summary see \cite{2TPhaseSpace}). Although the full machinery of
2T-physics is not needed for this cosmological application (only familiar
conformal symmetry in 3+1 dimensions is enough), it is important to emphasize
that the techniques and physical clues were first developed in the context of
2T-physics. The more powerful aspects of the 2T approach are also expected to
become relevant in a variety of applications in physics. For this reason I
give a brief introduction of 2T-physics in the following section, and then
concentrate on the relevant local scaling symmetry directly in 3+1 dimensions
in the sections after that.

\section{Basic Principles of 2T-Physics}

In this section I explain how the local scaling symmetry (Weyl symmetry
\cite{Weyl}) essential in our cosmological application emerges from a deep
principle of physics and how local scaling symmetry in cosmology is an example
of much broader hidden symmetries and dualities that are made manifest by
2T-physics as properties of a higher spacetime.

This section is included for two reasons. The first is to explain the
historical development of the ideas/methods, including the Weyl symmetry, that
ended up being useful in cosmology. The second, and more important reason is
to emphasize the richness of the underlying formulation that is waiting to be
applied to broad areas of physics. To understand the rest of the paper it is
not essential to digest the contents of this section (so the impatient reader
could skip to the next section).

2T-physics is a $4+2$ dimensional reformulation of physics in 3+1 dimensions.
It applies universally and correctly to all macroscopic or microscopic
phenomena at classical and quantum levels. It can be used to shed new light
on, and develop new computational methods in, well established fields of
physics, or it can be applied as a new guiding unification principle to
formulate and explore lesser known areas of physics, such as M-theory.

More generally, 2T-physics is formulated in $d$ space and $2$ time dimensions.
Note that in contrast to the 6 or 7 additional spatial extra dimensions in
string/M-theory, which usually are compactified, the extra 1+1 dimensions in
2T-physics are distinguished and not small. Thus, the 2T version of M-theory
would have a total of 13 dimensions. In fact, initially, 2T-physics emerged
from my observation that the extended supersymmetry of M-theory is actually a
SO$(10,2)$ covariant 12-dimensional supersymmetry written in disguise in an
11-dimensional basis (this 2-time signal in M-theory is distinct and prior to
F-theory which extended M-theory to 12 dimensions). This prompted the
development of S-theory in $11+2$ dimensions in 1996, which was an algebraic
unification of the supersymmetries of the various corners of M-theory within
the supergroup OSp$(1|64)$, which in turn led to 2T-physics after a few twists
and turns by 1998 (see \cite{2TPhaseSpace} for references on these
developments). A complete set of references and an elementary explanation of
2T-physics is given in \cite{ExtraDims}.

The elementary notions of 2T-physics in 1998, which were formulated in phase
space (position and momentum), evolved by now into a new principle of physics:
\textit{observer independence in phase space }\cite{2TPhaseSpace}. This
principle generalizes a similar notion championed by Albert Einstein who, with
his general coordinate symmetry, showed us how to formulate physics in a way
that the basic laws are independent of the status of observers is coordinate
space. The new principle takes this notion one step further into phase space,
since, after all, observers are defined with not only position but also
momentum. The stronger gauge symmetry at the base of 2T-physics, \textit{phase
space gauge symmetry,} goes beyond general covariance, or Yang-Mills type
gauge symmetries in position space, and explores gauge symmetry in phase space
that covers a much larger class of observers. As a consequence, 2T-physics
comes with a lot more gauge symmetry as compared to 1T-physics, and this
guides us to a more fundamental and unified form of the laws of physics, that
is independent of the bias of observers in phase space.

To be able to realize the phase space gauge symmetry, an extra space and an
extra time dimension must be added. The reason is that, in a spacetime with
only one time dimension, phase space gauge symmetry is too strong and reduces
the system to trivial content in the gauge invariant physical subspace. With
an extra 1-space and 1-time dimensions the desired gauge symmetry is
realizable and yields very rich content. Consequently, the difference between
2T-physics in $4+2$ dimensions and 1T-physics in 3+1 dimensions is only gauge
degrees of freedom. That is, there are no Kaluza-Klein degrees of freedom that
fill the gap between 3+1 and $4+2$. Instead, the available \textit{physical}
degrees of freedom (in the gauge invariant subspace) effectively describe 3+1
dimensional holographic \textquotedblleft shadows\textquotedblright\ of the
phenomena that have a more unified and more symmetric description in the full
$4+2$ dimensions.

To help grasp the relation between 1T-physics and 2T-physics, I suggested a
metaphor (see Fig.3 in \cite{2TPhaseSpace}). Consider the many possible
shadows of a 3 dimensional object projected from different perspectives on the
surrounding walls of a 3-dimensional room. A flatlander that can crawl and
measure only on the surface of the walls would think that the shadows of
different shapes are different \textquotedblleft beasts\textquotedblright\ and
move differently. Similarly, even though according to 2T-physics a unique
dynamical system in $4+2$ dimensions generates a large variety of 1-time
\textquotedblleft shadows\textquotedblright', 1T-physics presents these
\textquotedblleft shadows\textquotedblright\ in 3+1 dimensional space-times as
different dynamical systems in terms of different Hamiltonians (different
times). In this way 1T-physics misses the underlying relationship between the
\textquotedblleft shadows\textquotedblright\ as well as the underlying
properties (e.g. symmetries) of the higher dimensional space-time. Actually,
it turns out that each \textquotedblleft shadow\textquotedblright\ is a
holographic image that retains all the information of the d+2 structure. This
information takes the form of hidden symmetries, dualities and other
non-trivial structures, which are hard to notice by the 1T physicist that
investigates the \textquotedblleft shadows\textquotedblright\ (i.e. different
dynamical systems).

The advantage is that, the higher $4+2$ dimensions, that include the gauge
degrees of freedom, provide multiple new perspectives in higher spacetime
(observers in phase space) in which to view 3+1 dimensional phenomena with
more clarity and unity. This is somewhat analogous to Einstein's observers in
relativistic 4-dimensional spacetime whose very different views of physical
phenomena from their own perspectives in spacetime (defined by gauge choices)
are united by the underlying gauge symmetries that elevate time to the same
level of importance as space in the formulation of fundamental laws of
physics. Similarly, with phase space gauge symmetry, all components of
momentum and position (including time and Hamiltonian) are all at the same
level of importance in the formulation of fundamental laws of physics. In this
way, as in the case of Einstein's surprising conclusions, 2T-physics makes
numerous surprising predictions of previously unknown dualities and hidden
symmetries in 1T-physics, as natural consequences of the underlying phase
space gauge symmetry and global symmetry in $4+2$ dimensions. The $4+2$
dimensional equations unite diverse dynamical systems in 3+1 dimensions
(diverse Hamiltonians used by diverse 1T observers in 3+1 dimensional phase
space) and give them a single unified higher dimensional interpretation.

2T-physics is causal (no time-like closed loops or grandfather paradoxes), and
it is unitary (free of ghosts), due to the gauge symmetry, and describes
physics correctly at all levels of distance or energy. It has been developed
in the realms of classical or quantum particles with interactions, their
twistor equivalents, field theory including the Standard Model, gravity,
supergravity and high spin fields.

SO$\left(  4,2\right)  $ global symmetry, and its local extension, is the
natural spacetime symmetry of the ambient $4+2$ dimensional flat spacetime for
2T-physics (more generally d+2, if small compactified dimensions are included,
as in string theory). This global symmetry in $4+2$ dimensions takes the form
of conformal symmetry SO$\left(  4,2\right)  $ in 3+1 dimensions in the case
of massless systems, as well as the little known hidden symmetry of many other
systems, such as the Friedmann-Lema\^{\i}tre-Robertson-Walker (FLRW) universe,
the hydrogen atom, and numerous other cases (see Fig.1 in \cite{2TPhaseSpace}%
). This comes about as follows.

The 2T theory, in any of its forms mentioned above (particles, field theory,
twistors, etc.), may be gauge fixed from $4+2$ to 3+1 in many possible ways,
such that each gauge choice corresponds to the perspective of an observer in
3+1 dimensional phase space as embedded in $4+2$. It is evident that there are
an infinite number of ways in which 3+1 dimensional phase space sits within
$4+2$ dimensional phase space (the extra phase space in 1+1 dimensions is
gauge degrees of freedom). Each such perspective provides a \textquotedblleft
shadow\textquotedblright\ of the phenomena in $4+2$ dimensions as seen by an
observer using that 3+1 phase space perspective. All such observers are
related to each other by canonical transformations in 3+1 dimensions, where
such transformations are gauge transformations in the 2T-physics formalism.
Any global symmetry (such as SO$\left(  4,2\right)  $, or other global
symmetry in a specific model in $4+2$), that \textit{commutes with the gauge
symmetry}, remains unspoiled in all shadows. This is similar to the global
Lorentz symmetry that remains as a hidden symmetry in Maxwell's
electro-magnetism or any Yang-Mills theory when one chooses a gauge, such as
the Coulomb gauge. Thus, all shadows share the same global symmetry inherited
from $4+2$ dimensions, while all shadows are related to each other by
dualities since (being gauge choices) they all are holographic representations
of the same parent theory in $4+2$ dimensions.

The richness (and usefulness) of 2T-physics is in its ability to unify the
diverse forms of 1T-physics as perceived in various shadows and in
establishing previously unknown dualities and hidden symmetries that are
implemented via canonical transformations among shadows. This predicted
additional information that relates \textquotedblleft
shadows\textquotedblright\ to each other is missing systematically in
1T-physics (although some such cases could, and have been, accidentally
discovered within 1T-physics). The applications of this new systematic
concepts, for example, to develop new computational tools in 1T-physics, or to
gain deeper insight into spacetime and physics, are still largely underdeveloped.

There is a particular perspective (gauge choice) called the \textquotedblleft
conformal shadow\textquotedblright\ in which the linear SO$\left(  4,2\right)
$ Lorentz symmetry in flat $4+2$ dimensions is perceived by the corresponding
observer in 3+1 dimensions as the familiar \textit{conformal symmetry} of 3+1
dimensions. Most work in 2T-physics has concentrated on the \textit{conformal
shadow} because it is then possible to connect to the familiar formalism of
relativistic field theory used in particle physics and gravity. Then it is
possible to interpret the familiar conformal symmetry as one perspective in
which to view the phenomena in $4+2$ dimensions. Although this is a very
useful perspective to connect to familiar language, other perspectives would
provide additional information about the underlying unity in the form of dual
versions of the same theory that may be used both as computational tools as
well as for deeper insights into spacetime and the corresponding physics.

The cosmological application discussed in this paper emerged from the analysis
of 2T-gravity \cite{2Tgravity} in the conformal shadow
\cite{2TgravityGeometry}. A necessary consequence of 2T-gravity is that the
emergent 1T-gravity in 3+1 dimensions must come with an additional local
scaling symmetry, known as Weyl symmetry, realized in a specific form. This is
possible but not required in 1T field theory. This local scaling symmetry is a
remnant of general coordinate reparameterization symmetry in $4+2$ dimensions;
specifically, it is reparameterizations of the extra 1+1 dimensions locally at
each point of the 3+1 dimensions \cite{2TgravityGeometry}. To the observer in
the conformal shadow, the Weyl symmetry must be present as an essential
property of gravity and all the degrees of freedom that couple to it. Making a
gauge choice for the Weyl symmetry amounts to \textit{choosing the observer's
coordinates in the extra 1-space and 1-time dimensions}.

To summarize, the connection of 2T-physics to conformal symmetry is dictated
by the phase space gauge symmetry. This underlying gauge symmetry requires
$4+2$ dimensions (more gerally $d+2).$ The spacetime symmetries in $4+2$
dimensions that commute with the gauge symmetry, such as the linear SO$\left(
4,2\right)  $ in \textit{flat} $4+2$ dimensional spacetime, remains as a
hidden symmetry in all shadows. In the particular shadow called the conformal
shadow, the hidden SO$\left(  4,2\right)  $ takes the form of the non-linearly
realized conformal SO$\left(  4,2\right)  $ transformations in 3+1 dimensions.
In \textit{curved} $4+2$ spacetime the symmetry includes general coordinate
transformations of all $4+2$ coordinates. In the conformal shadow, at each 3+1
dimensional point, the reparametrizations of the extra 1+1 dimensions take the
form of local scaling transformations, which is the Weyl transformation. Thus,
2T-physics requires that all 3+1 dimensional physics must be formulated in a
Weyl symmetric formalism, whose form is already predicted by the structure of
the $4+2$ dimensional approach.

From this point on, using only 3+1 dimensional field theory methods, and
requiring the Weyl symmetry, is sufficient to proceed without further guidance
from 2T-physics. The Weyl symmetry plays multiple roles: (i) it restricts the
terms that are allowed in the gravitational action and matter couplings, (ii)
it is used as a tool for solving the well known cosmological equations of
Einstein's theory by taking advantage of various gauge choices (a super
simplified example of the observers, shadows and dualities mentioned above),
(iii) it is a guide for completing the geometry geodesically.

\section{Using the Weyl Symmetry to Solve Cosmological Equations}

The Weyl symmetry does not allow any dimensionful parameters in the action
$S=\int d^{4}x\sqrt{-g}\mathcal{L}\left(  x\right)  $ that describes the
coupling of gravity and matter fields. Here $S$ is the action, $\mathcal{L}%
\left(  x\right)  $ the Lagrangian density, and $g$ the determinant of the
metric $g_{\mu\nu}\left(  x\right)  .$ In particular, the usual
Einstein-Hilbert term is not permitted since it contains a dimensionful
parameter. However, what is permitted is \textit{conformally coupled scalars}
\cite{deser} as well as the usual gravitational coupling to fermions and gauge
fields. A full theory that includes the Standard Model or its supersymmetric
or grand unified extensions, and their coupling to gravity, including a mass
generating mechanism for the Higgs field consistent with Weyl symmetry, can be
constructed. Indeed this was done directly in $4+2$ dimensions
\cite{2tStandardModel}\cite{2tSM2}, and its conformal shadow is precisely the
emergent Weyl invariant theory in 3+1 dimensions.

For the purpose of discussing cosmology we concentrate only on two conformally
coupled scalar fields, $\phi,s$, for which the \textit{Weyl invariant} action
takes the following form \cite{2Tgravity}\cite{2TgravityGeometry}
\begin{equation}
\mathcal{L}\left(  x\right)  =\frac{1}{12}\left(  \phi^{2}-s^{2}\right)
R\left(  g\right)  +g^{\mu\nu}\left(  \frac{1}{2}\partial_{\mu}\phi
\partial_{\nu}\phi-\frac{1}{2}\partial_{\mu}s\partial_{\nu}s\right)  -\phi
^{4}f\left(  s/\phi\right)  +radiation+matter. \label{action1}%
\end{equation}
where $R\left(  g\right)  $ is the curvature for the metric $g_{\mu\nu},$ and
$f\left(  s/\phi\right)  $ is an arbitrary function of its argument $s/\phi.$
This action is invariant under the local Weyl symmetry%
\[
\phi\rightarrow\Omega\phi,\;s\rightarrow\Omega s,\;g_{\mu\nu}\rightarrow
\Omega^{-2}g_{\mu\nu},
\]
where $\Omega\left(  x\right)  $ is an arbitrary function of spacetime. Things
to notice include:

\begin{itemize}
\item There is no Einstein-Hilbert term $R\left(  g\right)  /(16\pi G),$ where
$G$ is the Newton constant, because this is not invariant under the Weyl
symmetry. Instead, the factor $\left(  \phi^{2}-s^{2}\right)  /12$ that
multiplies $R\left(  g\right)  $ will generate the Newton constant by choosing
a particular gauge for the Weyl symmetry, which we call the Einstein gauge.

\item The field $\phi$ has the wrong sign kinetic term so it is potentially a
ghost; but since there is a Weyl symmetry that compensates for it, the theory
is unitary and there is no ghost.

\item There can be any number of scalar fields, $s_{1},s_{2},\cdots,$ that
appear just like $s,$ all of which are conformally coupled to preserve the
Weyl symmetry \cite{2tSugra}\footnote{There are more general ways to couple
scalars to gravity consistent with Weyl symmetry as given in \cite{2tSugra}.
In that case the factor $\left(  \phi^{2}-s^{2}\right)  /12$ is replaced by a
more general function $U(\phi,s_{i})$ where $U$ obeys specific constraints.
But this also requires non-canonical kinetic terms for the scalars
(sigma-model-like couplings which are derived from $U$). If one requires only
canonically normalized scalars, then the only chopice is $U=\left(  \phi
^{2}-\sum_{i}s_{i}^{2}\right)  /12.$}. All such scalars must have the correct
sign of the kinetic energy, because there is not enough gauge symmetry to
compensate for ghosts other than the field $\phi.$ Then the $s^{2}$ in the
factor $\left(  \phi^{2}-s^{2}\right)  /12$ is replaced by $s^{2}=s_{1}%
^{2}+s_{2}^{2}+\cdots.$ One of these scalars could be identified with the
dilaton in string theory as in \cite{BCST2}, a subset that forms a doublet
under the electroweak gauge group SU$\left(  2\right)  \times$U$\left(
1\right)  $ may be identified with the Higgs field, and so on \cite{2Tgravity}.

\item Changing the sign of the kinetic energy for $\phi$ would lead to the
purely negative factor $\left(  -\phi^{2}-s^{2}\right)  /12$ multiplying $R$
in order to preserve the Weyl symmetry. Then there can be no spacetime patch
in which the effective gravitational constant could be positive. Hence there
must be one scalar with the wrong sign kinetic energy, namely $\phi,$ so that
the factor $\left(  \phi^{2}-s^{2}\right)  /12$ that multiplies $R\left(
g\right)  $ can be positive at least in some patches of spacetime $x^{\mu}.$
In a supersymmetric version of this theory all scalars must be complexified.
This requires more gauge symmetry beyond Weyl symmetry to remove the ghost for
the complex $\phi.$ Such a local symmetry exists as a hidden symmetry in
supergravity as discussed in \cite{2tSugra}, so this formalism has been
extended to the supersymmetric version as well.
\end{itemize}

This form of action in 3+1 dimensions emerged from 2T-gravity in $4+2$
dimensions \cite{2Tgravity}\cite{2TgravityGeometry}. A natural question in
2008 was: can the factor $\left(  \phi^{2}-s^{2}\right)  /12$ change sign
dynamically even if one starts with positive initial conditions $\left(
\phi_{0}^{2}-s_{0}^{2}\right)  >0$ in some patch of spacetime? If the answer
is no, then a positive gauge choice $\left(  \phi^{2}-s^{2}\right)
/12\rightarrow(16\pi G)^{-1}$ would be possible for all spacetime $x^{\mu}$,
and gravity everywhere for all time would be experienced as described by the
Einstein-Hilbert term. But if the answer is yes, it would imply that the
universe could include patches of spacetime in which $\left(  \phi^{2}%
-s^{2}\right)  /12$ can be gauge fixed to a negative constant; in those
regions of spacetime there would be antigravity, namely, a negative Newton
constant, therefore a repulsive gravitational force.

In spacetime patches where $\left(  \phi^{2}-s^{2}\right)  \geq0,$ the theory
in Eq.(\ref{action1}) is equivalent to the Einstein-Hilbert version of
gravity. This is shown by choosing the Einstein gauge in which we label all
fields with an extra index $E,$ such as $\phi_{E},s_{E},g_{E}^{\mu\nu},$ to
indicate they are gauge fixed so that $\left(  \phi_{E}^{2}-s_{E}^{2}\right)
/12=(16\pi G)^{-1}$ is a constant for all $x^{\mu}$ in the patch. This is done
by parameterizing $\phi_{E},s_{E}$ in terms of a field $\sigma\left(
x\right)  $ as
\begin{equation}
\phi_{E}\left(  x\right)  =\pm\frac{\sqrt{12}}{\sqrt{16\pi G}}\cosh\left(
\frac{\sqrt{16\pi G}}{\sqrt{12}}\sigma\left(  x\right)  \right)
,\;s_{E}\left(  x\right)  =\pm\frac{\sqrt{12}}{\sqrt{16\pi G}}\sinh\left(
\frac{\sqrt{16\pi G}}{\sqrt{12}}\sigma\left(  x\right)  \right)  ,
\label{Egauge}%
\end{equation}
Inserting this in Eq.(\ref{action1}) the gauge fixed form of the action is
obtained
\begin{equation}
S=\int d^{4}x\sqrt{-g_{E}}\left\{  \frac{1}{16\pi G}R\left(  g_{E}\right)
-\frac{1}{2}g_{E}^{\mu\nu}\partial_{\mu}\sigma\partial_{\nu}\sigma-V\left(
\sigma\right)  +radiation+matter\right\}  . \label{action gravity}%
\end{equation}
where the arbitrary potential $V\left(  \sigma\right)  $ is related to the
arbitrary function $f\left(  s/\phi\right)  .$ Hence, the cosmological
equations of the standard scalar-tensor theory in Eq.(\ref{action gravity})
are the same as those or the Weyl invariant action in any spacetime patch in
which $\left(  \phi^{2}-s^{2}\right)  \geq0.$

The patches in which $\left(  \phi^{2}-s^{2}\right)  $ is negative provide an
extension of the domain of the field space of Einstein's theory. The full
field space in $\phi,s$ plane and the regions of gravity/antigravity are shown
in Fig.1.
\begin{figure}
[h]
\begin{center}
\includegraphics[
height=2.1854in,
width=2.1854in
]%
{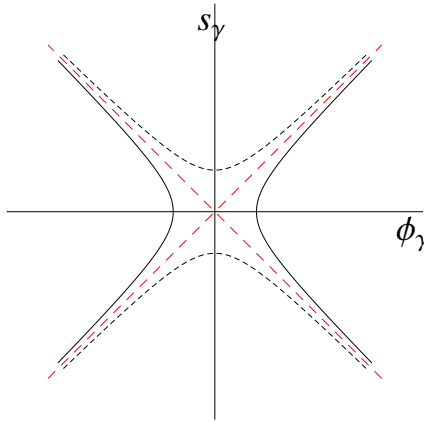}%
\caption{The $\left(  \phi,s\right)  $ plane for the scalar fields $\phi,s,$
taken in the $\gamma$-gauge defined in Eq.(\ref{gaugeGamma}). The dotted
diagonal lines is where $\left(  1-s^{2}/\phi^{2}\right)  \left(  x^{\mu
}\right)  =0$ in all gauges. With respect to the diagonals, the left and right
quadrants is the gravity region where $\left(  \phi^{2}-s^{2}\right)  \left(
x^{\mu}\right)  >0$ in all gauges, and the top and bottom quadrants is the
antigravity region where $\left(  \phi^{2}-s^{2}\right)  \left(  x^{\mu
}\right)  <0$ in all gauges. The geodesically incomplete Einstein-Hilbert
formulation of gravity resides in the right + left quadrants. The geodesically
complete theory is obtained by extending the domain of the fields $\left(
\phi,s\right)  $ to include the full $\left(  \phi,s\right)  $ plane. The
diagonals is where the cosmological singularity Crunch or Bang occurs
\textit{in the Einstein frame} as seen from Eq.(\ref{E=gamma}). The origin of
the $\left(  \phi,s\right)  $ plane, $\phi=s=0,$ which is only one point on
the diagonals, is the only Crunch-Bang transition point permitted by the
attractor mechanism.}%
\label{fig1}%
\end{center}
\end{figure}
When $\left(  \phi^{2}-s^{2}\right)  \leq0,$ that is in the antigravity
regime, it is again possible to choose an Einstein gauge in which $\left(
\phi_{E}^{2}-s_{E}^{2}\right)  /12=-(16\pi G)^{-1},$ so the gravitational
constant is negative. This amounts to switching $\cosh$ and $\sinh$ in
Eq.(\ref{Egauge}), and consequently changing the signs of the first two terms
(but not the other signs!) in the gauge fixed action in
Eq.(\ref{action gravity}). Thus the Weyl invariant action in Eq.(\ref{action1}%
) includes an antigravity sector in the traditional language of the Einstein
frame. The two sectors are related by simply extending the domain of the
fields $\left(  \phi,s\right)  $ to cover the entire $\left(  \phi,s\right)  $
plane in Fig.1, rather than only the gravity regime defined by the restriction
$\left(  \phi^{2}-s^{2}\right)  \geq0.$ The question was: are the
gravity/antigravity sectors connected by dynamics?

Informal conversations over two years with numerous cosmologists around the
world did not produce an answer to the question of whether the dynamics
compels $\left(  \phi^{2}-s^{2}\right)  $ to change sign. So an
\textit{analytic} study of the dynamical equations needed to be undertaken, as
we did in 2010 \cite{inflation-BC}. Playing with various gauge choices of the
Weyl symmetry, produced tricks and methods to solve the cosmological equations
analytically for some choices of the potential energy described by the
function $\phi^{4}f\left(  s/\phi\right)  $ or $V\left(  \sigma\right)  .$
Through the analytic solutions, it became apparent that the sign of the gauge
invariant quantity $\left(  1-s^{2}/\phi^{2}\right)  \left(  x^{\mu}\right)  $
can change dynamically, at least as a function of time. Note that the sign of
$\left(  1-s^{2}/\phi^{2}\right)  $ or equivalently of $\left(  \phi^{2}%
-s^{2}\right)  $ is a Weyl gauge invariant. We expected that it would be hard
to find solutions of the equations in which the sign changed dynamically, but
instead we discovered that the change of sign happens generically. To prevent
the sign switch from happening one must artificially fine tune initial
conditions of the fields $\phi,s,g_{\mu\nu},$ and/or parameters in the
potential energy $V\left(  \sigma\right)  $ or $\phi^{4}f\left(
s/\phi\right)  .$ We also understood that, the spacetime point(s) at which the
gauge invariant factor $\left(  1-s^{2}/\phi^{2}\right)  \left(  x^{\mu
}\right)  =0$ vanishes just before changing sign in \textit{any gauge}, is a
curvature singularity of $R(g_{E})$ in the \textit{Einstein gauge}. Thus, in
the $E$-gauge in which $\left(  \phi_{E}^{2}-s_{E}^{2}\right)  $ has been
gauge fixed to a constant, the gauge invariant $\left(  1-s_{E}^{2}/\phi
_{E}^{2}\right)  $ vanishes exactly at the curvature singularity, while
$\phi_{E}^{2}$ goes to infinity to maintain the constant gauge.

Thus the sign change is a generic consequence of dynamics, and the sign can
change only at a geometrical singularity. If the theory of Eq.(\ref{action1})
is restricted by hand to the sector $\left(  \phi^{2}-s^{2}\right)  \left(
x^{\mu}\right)  \geq0$ then the spacetime $\left\{  x^{\mu}\right\}  $ will be
cutoff at the singularity, and the theory will be geodesically incomplete.
This explained precisely how the traditional formulation of gravity in the
form of Eq.(\ref{action gravity}) is geodesically incomplete, since it amounts
to only the geodesically incomplete positive domain $\left(  \phi^{2}%
-s^{2}\right)  \left(  x^{\mu}\right)  \geq0$. The cure to make it
geodesically complete is evident: extend the Einstein frame with the Weyl
symmetry (i.e. reintroduce the gauge degrees of freedom $\phi,s$ rather than
just $\sigma$) and add the missing antigravity domain shown in Fig.1. This
extension of the domain in field space is reminiscent of the Kruskal-Szekeres
extension of the domain of the spacetime metric across the horizon of the
blackhole to include the interior region.

Since singularities were involved with the switching of signs, a collaboration
with cosmologists ensued \cite{BCT1}-\cite{BCST2} to try to understand more
thoroughly the cosmological significance of the switch of the sign of $\left(
\phi^{2}-s^{2}\right)  $. The fact that the action in Eq.(\ref{action1}) could
be directly related to the colliding branes scenario of a cyclic universe
\cite{cyclic1}\cite{branesMT}\footnote{This is seen after massaging some more
the equations in \cite{branesMT}. This unpublished manipulation of the
equations that made the connection more evident was developed in the process
of the collaboration in \cite{BCT1}.} made it even more attractive to study
the model in Eq.(\ref{action1}) for cosmological applications.

The analytic solutions were obtained easily in another gauge called the
$\gamma$-gauge in which the fields are labelled with the letter $\gamma,$ such
as $\phi_{\gamma},s_{\gamma},g_{\gamma}^{\mu\nu},$ to indicate they are gauge
fixed differently than some other useful gauges. The $\gamma$-gauge is defined
by demanding that the determinant of the metric is fixed to $1,$ such as
\begin{equation}
\det(g_{\gamma}\left(  x\right)  )=-1 , \label{gaugeGamma}%
\end{equation}
for all spacetime $x^{\mu}.$ We define the following Weyl gauge invariant
quantity $\chi$ which plays an important role in our discussion%
\begin{equation}
\chi\equiv\frac{16\pi G}{12}(-g)^{\frac{1}{4}}(\phi^{2}-s^{2}). \label{chi}%
\end{equation}
Other useful gauge invariants are the ratio $s/\phi$ and the products
$(-g)^{\frac{1}{8}}\phi$ and $(-g)^{\frac{1}{8}}s$. By gauge invariance we can
equate the gauge fixed forms of $\chi$ in the $E$-gauge and in the $\gamma
$-gauge, which yields (for all signs of $(\phi^{2}-s^{2})$)%
\begin{equation}
(-g_{E})^{\frac{1}{4}}=\frac{16\pi G}{12}\left\vert \phi_{\gamma}%
^{2}-s_{\gamma}^{2}\right\vert . \label{E=gamma}%
\end{equation}
This shows that when $(\phi_{\gamma}^{2}-s_{\gamma}^{2})$ vanishes in the
$\gamma$-gauge, the gauge invariant quantity $\left(  1-s^{2}/\phi^{2}\right)
$ will vanish in all gauges, while \textit{only} in the $E$-gauge the
determinant of the metric $(-g_{E})$ will also vanish. This shows that the
complete \textit{failure of the geometry in the Einstein gauge} (which is when
the cosmological singularity occurs, and geodesics are incomplete) is related
directly to the vanishing of the factor $(\phi^{2}-s^{2})$ that appears in the
Weyl invariant action.

Note that the failure of the geometry in the Einstein gauge does not imply the
failure of the geometry in other Weyl gauges. Recall that curvature does
transform under Weyl transformations. In particular, by construction, since
$\det(g_{\gamma})=-1$ in all spacetime patches in the $\gamma$-gauge, we may
expect smoother behavior of curvature components $R_{\mu\nu\lambda\sigma
}\left(  g_{\gamma}\right)  $ in the $\gamma$-gauge (even if singular for some
components), as compared to more singular curvature components in the Einstein
gauge $R_{\mu\nu\lambda\sigma}\left(  g_{E}\right)  $. This smoother feature
of the $\gamma$-gauge, and some additional properties of some potential energy
functions $V\left(  \sigma\right)  $ or $\phi^{4}f\left(  s/\phi\right)  ,$
turned out to be the key factors both to \textit{geodesically complete} the
geometry of the Einstein frame, and to \textit{find analytically the complete
set of solutions} of all the fields, in a homogeneous cosmological state (i.e.
purely time dependent fields).

For the purpose of cosmological studies, we concentrate on the relevant
geometries of interest that can be written in the form $ds^{2}=a^{2}\left(
\tau\right)  \left(  -e^{2}d\tau^{2}-ds_{3}^{2}\right) ,$ where $\tau$ is the
conformal time, $a\left(  \tau\right)  $ is the scale factor of the universe,
$e\left(  \tau\right)  $ is the lapse function that plays the role of a gauge
field that insures $\tau$-reparameterization symmetry, and $ds_{3}^{2}$ is the
spacial metric in 3-dimensions in which we will include a curvature parameter
$k$ and anisotropy degrees of freedom as functions of time $\alpha_{1}\left(
\tau\right)  ,\alpha_{2}\left(  \tau\right)  .$ We consider the spaces known
as FRLW, Kasner, Bianchi IX and Bianchi VIII. All of these are captured by the
following parametrization of the 3-dimensional part of the metric (the
$\alpha_{1,2}$ used in this paper are redefined as compared to \cite{BCST1}%
\cite{BCST2} by rescaling them with the factor $\sqrt{16\pi G/12},$ so that
here they are dimensionless)
\begin{equation}
\left(  ds_{3}^{2}\right)  =e^{-4\alpha_{1}\left(  \tau\right)  }\left(
d\sigma_{z}\right)  ^{2}+e^{2\alpha_{1}\left(  \tau\right)  }\left(
e^{2\sqrt{3}\alpha_{2}\left(  \tau\right)  }\left(  d\sigma_{x}\right)
^{2}+e^{-2\sqrt{3}\alpha_{2}\left(  \tau\right)  }\left(  d\sigma_{y}\right)
^{2}\right)  . \label{ds3}%
\end{equation}
For FRLW that has no anisotropy we set $\alpha_{1,2}\rightarrow0$ and $\left(
ds_{3}^{2}\right)  _{FRLW}=\left(  dx^{2}+dy^{2}+dz^{2}\right)  /(1-k\left(
x^{2}+y^{2}+z^{2}\right)  ^{2}).$ For Kasner that has no curvature we set
$\left(  d\sigma_{x},d\sigma_{y},d\sigma_{z}\right)  =\left(  dx,dy,dz\right)
,$ and for the Bianchi metrics that have both anisotropy and curvature
$\left(  d\sigma_{x},d\sigma_{y},d\sigma_{z}\right)  $ \ include the curvature
parameter $k$ along with $\left(  x,y,z\right)  $ dependence.

Then concentrating on purely time dependent homogeneous fields, the effective
action relevant for cosmological investigations that follows from
Eq.(\ref{action1}) is%
\begin{equation}
S_{\text{eff}}=\int d\tau\left\{
\begin{array}
[c]{c}%
-\frac{1}{2e}\left(  \partial_{\tau}\left(  a\phi\right)  \right)  ^{2}%
+\frac{1}{2e}\left(  \partial_{\tau}\left(  as\right)  \right)  ^{2}+\frac
{1}{2e}\left(  \phi^{2}-s^{2}\right)  a^{2}\left(  \dot{\alpha}_{1}^{2}%
+\dot{\alpha}_{2}^{2}\right) \\
-e\left[  a^{4}\phi^{4}f\left(  \frac{s}{\phi}\right)  +\rho_{r}-\frac{1}%
{2}\left(  \phi^{2}-s^{2}\right)  a^{2}v\left(  \alpha_{1},\alpha_{2}\right)
\right]  ,
\end{array}
\right\}  , \label{SeffConf}%
\end{equation}
where $\rho_{r}$ is the radiation density when the scale factor $a=1$. Here
$v\left(  \alpha_{1},\alpha_{2}\right)  $ is the anisotropy potential which
emerges from the curvature term $\left(  \phi^{2}-s^{2}\right)  R\left(
g\right)  $
\begin{equation}
v\left(  \alpha_{1},\alpha_{2}\right)  =\frac{k}{1-4~\text{sign}\left(
k\right)  }\left[
\begin{array}
[c]{c}%
e^{-8\alpha_{1}}+4e^{4\alpha_{1}}\sinh^{2}\left(  2\sqrt{3}\alpha_{2}\right)
-4~\text{sign}\left(  k\right)  ~e^{-2\alpha_{1}}\cosh\left(  2\sqrt{3}%
\alpha_{2}\right)
\end{array}
\right]  . \label{Va}%
\end{equation}
In the isotropic limit the anisotropy potential reduces to a constant
$\lim_{\alpha_{1,2}\rightarrow0}v\left(  \alpha_{1},\alpha_{2}\right)  =k.$
For the Kasner metric the potential energy term is absent even if anisotropy
is present since $k=0.$

The homogeneous degrees of freedom in $S_{\text{eff}}$ are the scalar fields
$\left(  \phi\left(  \tau\right)  ,s\left(  \tau\right)  \right)  $ and the
geometric fields $\left(  a\left(  \tau\right)  ,\alpha_{1}\left(
\tau\right)  ,\alpha_{2}\left(  \tau\right)  ,e\left(  \tau\right)  \right)  $
that are part of the metric. This effective action is invariant under the
gauge symmetries of $\tau$-reparameterization and $\tau$-dependent Weyl
transformations
\begin{equation}
a\left(  \tau\right)  \rightarrow\Omega^{-1}\left(  \tau\right)  a\left(
\tau\right)  ,\;\left(  \phi\left(  \tau\right)  ,s\left(  \tau\right)
\right)  \rightarrow\Omega\left(  \tau\right)  \left(  \phi\left(
\tau\right)  ,s\left(  \tau\right)  \right)  ,
\end{equation}
while $\alpha_{1,2}$ and $e$ are Weyl invariant. After using the equations of
motion for $e$ (that results in the zero energy constraint), we choose the
gauge $e=1.$ Furthermore, one of the fields $\left(  a,\phi,s\right)  $ may be
eliminated by a Weyl gauge choice; for each choice the fields $\left(
a,\phi,s\right)  $ are labelled with an extra index to indicate that they are
the gauge fixed versions of the fields. We have discussed several gauge
choices (for a subset see \cite{BCST2}), including the Einstein gauge of
Eq.(\ref{Egauge}) labelled by $E$ as $\left(  a_{E},\phi_{E},s_{E}\right)  ,$
the $\gamma$-gauge of Eq.(\ref{gaugeGamma}) labeled by $\gamma$ as $\left(
a_{\gamma},\phi_{\gamma},s_{\gamma}\right)  $ in which $a_{\gamma}\left(
x\right)  =1$, the constant (or supergravity) gauge \cite{2Tgravity} labelled
by $c$ in which $\phi_{c}\left(  x\right)  =$constant$,$ and the string gauge
labelled by $s$ \cite{BCST2} in which the action in Eq.(\ref{action1}) reduces
to the form of the low energy effective action of string theory including the
dilaton in the string frame. The $E$-gauge is useful to discuss the physics in
the Einstein frame where most of the physical intuition on gravitational
physics has been developed historically. However, the cosmological field
equations are non-linear and difficult to solve in the $E$-gauge. On the other
hand, in the $\gamma$-gauge the field equations simplify greatly to the point
where they can be solved analytically to yield the full set of solutions for
some cases of the potential energy $V\left(  \sigma\right)  $ or $\phi
^{4}f\left(  s/\phi\right)  .$ This was exploited in \cite{inflation-BC}%
-\cite{BCST2} to obtain the results summarized in this paper.

The $E$-gauge and $\gamma$-gauge degrees of freedom are related by a gauge
transformation as in Eqs.(\ref{Egauge},\ref{E=gamma}) and
\begin{equation}
\text{ }a_{E}^{2}={\frac{16\pi G}{12}}\left\vert \phi_{\gamma}^{2}-s_{\gamma
}^{2}\right\vert =\left\vert \chi\right\vert ,\;\sigma=\frac{1}{2}\sqrt
{\frac{12}{16\pi G}}\ln\left\vert \frac{\phi_{\gamma}+s_{\gamma}}{\phi
_{\gamma}-s_{\gamma}}\right\vert . \label{link1}%
\end{equation}
This was used to obtain the solutions for $a_{E}\left(  \tau\right)  $ and
$\sigma\left(  \tau\right)  $ in the Einstein frame from the solutions for
$\phi_{\gamma}\left(  \tau\right)  ,s_{\gamma}\left(  \tau\right)  $ in the
$\gamma$-gauge. Similarly transformations among other gauges $E,\gamma,c,s$
leads to the solutions in other gauges.

The methodical study in \cite{inflation-BC}-\cite{BCST2} using
analytic tools will not be repeated here. Instead, only some results
will be outlined briefly. The complete set of solutions of the
cosmological equations were obtained with \textit{no restrictions on
either the parameters of the model or the initial conditions on the
fields}. Before the important role of anisotropy is taken into
account near the singularity at $a_{E}^{2}=0$, there are 6
unrestricted parameters in our solutions. All 6 parameters are
available to try to fit cosmological data far away from the
singularity. In the absence of anisotropy, the solution for $\left(
\phi_{\gamma}\left(  \tau\right) ,s_{\gamma}\left(  \tau\right)
\right)  $ behaves in various detailed ways in different regions of
the 6 parameter space. This is given in the Appendix of
\cite{BCST2}, where 25 different regions of the 6 parameter space
are identified in which the analytic expression is different for
each case separately. The trajectory of the solution in the absence
of anisotropy can be plotted parametrically (by eliminating the
$\tau$ parameter) in the $\left( \phi_{\gamma},s_{\gamma}\right)  $
plane of Fig.1. This shows that the \textit{generic trajectory}
crosses the lightcone in Fig.1 at any point so that $\left(
\phi_{\gamma}^{2}\left(  \tau\right)  -s_{\gamma}^{2}\left(
\tau\right)  \right)  $ keeps changing sign as the universe moves
between gravity/antigravity patches. To do so, the scale factor
$a_{E}^{2}\left( \tau\right)  $ vanishes at each crossing (according
to Eq.(\ref{link1})), so that the gravity/antigravity patches are
connected to each other only through the cosmological singularity in
the Einstein frame. It should be emphasized that the cosmological
singularity in the $E$-gauge at $a_{E}^{2}\rightarrow0$ is not a
singularity of the determinant of the $\gamma$-gauge metric, since
$a_{\gamma}=1$. The singularity of the $E$-gauge, which comes from
the vanishing of the gauge invariant $\chi\rightarrow0,$ becomes
$(\phi_{\gamma }^{2}-s_{\gamma}^{2})\rightarrow0$ in the
$\gamma$-gauge. However, within the $\gamma$-gauge, in the absence
of anisotropy, this is not at all singularity of the equations, and
in the presence of anisotropy  it is avoidable by several symmetry
arguments and a complex continuation \cite{BCST1}; this is why it
was possible to integrate the equations unambiguously as
$(\phi_{\gamma}^{2}-s_{\gamma}^{2})\rightarrow0,$ and understand the
geodesic completeness of the cosmological system.

An example of the generic solution in the \textit{absence} of anisotropy is
plotted in Fig.2a. At each crossing of the dotted diagonal lines, when
$a_{E}^{2}\left(  \tau_{i}\right)  =0$ at times $\tau_{i},$ the universe has a
zero-size bounce, where it transits \textit{generically} between
gravity/antigravity regions. Although the analytic solutions were obtained for
specific potentials, it became evident that this general behavior is generic
and basically model independent.

The behavior of the generic solution near the singularity changes dramatically
in the \textit{presence} of anisotropy by an attractor mechanism. An important
effect of anisotropy is to focus the trajectory of the \textit{generic
solution} to pass through the origin of the $\left(  \phi_{\gamma},s_{\gamma
}\right)  $ plane, such that the generic trajectory can cross the lightcone in
Fig.2b only at the origin where $\left(  \phi_{\gamma},s_{\gamma}\right)  $
vanish simultaneously.%

\begin{figure}
[tbh]
\begin{center}
\includegraphics[
height=3.3624in,
width=2.6671in
]%
{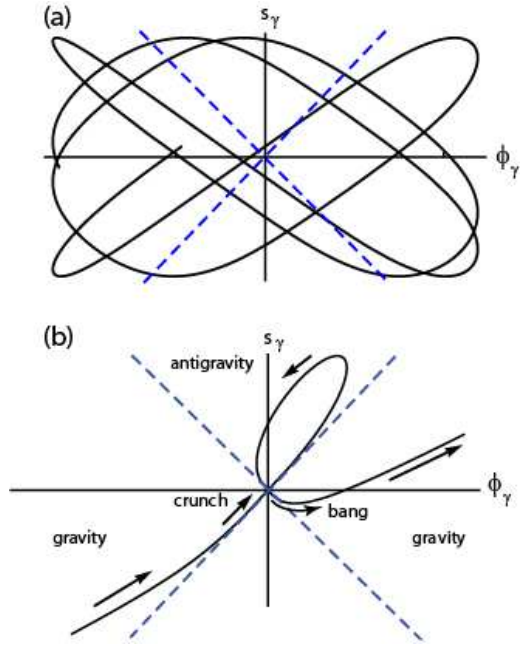}%
\caption{Comparison of the generic solution without anisotropy and
with anisotropy. (a) Trajectories in the $\left(
\phi_{\gamma},s_{\gamma}\right) $ plane for the generic solution,
with no anisotropy, cross the diagonal dotted lines at any point
that satisfies $\left\vert \phi_{\gamma}\right\vert =\left\vert
s_{\gamma}\right\vert $ (as determined by 6 parameters) where
$a_{E}^{2}$ vanishes. The crossing points is when the universe goes
through the crunch/bang, transiting between the gravity/antigravity
regions. (b) The generic solution with anisotropy is shown in the
vicinity of the singularity. An attractor mechanism caused by the
anisotropy distorts the path that would have crossed the dotted
diagonals (where $a_{E}^{2}=0$), so that it cannot cross them except
at the origin. Thus the universe passes only through
$\phi_{\gamma}=s_{\gamma}=0$ at the crunch, undergoes a loop in the
antigravity region, passes through the
origin again, and then re-emerges in the gravity regime. }%
\end{center}
\end{figure}

%

\begin{figure}
[tbh]
\begin{center}
\includegraphics[
height=1.9138in,
width=3.403in
]%
{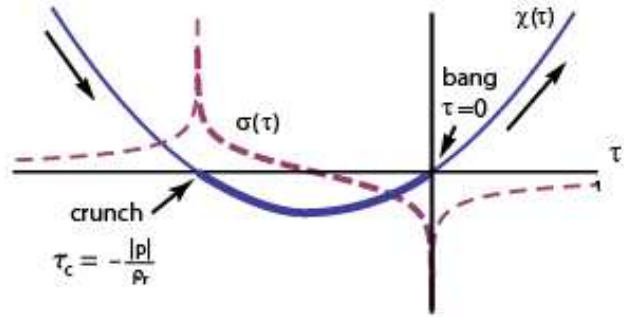}%
\caption{Plots of the gauge invariants $\chi(\tau)$ and
$\sigma(\tau)$ given in Eqs.(\ref{chi},\ref{link1}). The instants
$\tau=\tau_{c}$ and $\tau=0$ correspond to the big crunch and big
bang respectively. At these points the scale factor in the Einstein
frame $a_{E}$ vanishes. The big crunch/big bang transition is
punctuated by a brief period of antigravity between the times
$\tau=\tau_{c}$ and $\tau=0$ (thick portions of the curves) during
which $\left(  \phi^{2}-s^{2}\right)  $ is negative. The amount of
time spent in antigravity, namely $\vert \tau_{c}\vert\equiv
\frac{\vert{p}\vert}{\rho_r}$ where
$\vert{p}\vert=\sqrt{p_{\sigma}^{2}+p_{1}^{2}+p_{2}^{2}},$ is
determined by radiation $\rho_r$ and the canonical momenta $\left(
p_{\sigma},p_{1},p_{2}\right)  $ of the
$\sigma,\alpha_{1},\alpha_{2}$ fields, which are conserved during
the transition since the potential energy terms in the action
Eq.(\ref{SeffConf}) or Eq.(\ref{action1},\ref{action gravity}) are
negligible. }%
\end{center}
\end{figure}

The attractor mechanism discovered in \cite{BCST1} is \textit{model
independent} because the potential energy $V\left(  \sigma\right)  $ or
$\phi^{4}f\left(  s/\phi\right)  $ is negligible close to the singularity. It
is well known that the dominant terms in the Friedmann equations in the
Einstein frame are the kinetic energies for the scalar field and for
anisotropy. The next to the leading term is radiation represented by the
parameter $\rho_{r}$ in the action above. The potential energies for the
scalar field and anisotropy, which appear in the action in the form
$[a^{4}\phi^{4}f\left(  s/\phi\right)  -\frac{1}{2}\left(  \phi^{2}%
-s^{2}\right)  a^{2}v\left(  \alpha_{1},\alpha_{2}\right)  ] $ are negligible
as compared to the dominant terms, since, near the singularity, $a_{E}%
\rightarrow0$ in the $E$-gauge, or $\left(  \phi_{\gamma}^{2}-s_{\gamma}%
^{2}\right)  \rightarrow0$ in the $\gamma$-gauge. Through our
analytic analysis, we found that the mixmaster behavior described in
\cite{Misner} is avoided when the kinetic energy of the scalar field
is a few times larger than the kinetic energy of the anisotropy
fields as defined in the action of
Eq.(\ref{action1}); this is consistent with the discussion in \cite{BKL}%
\cite{Damour}. Hence the attractor solution given analytically in \cite{BCST1}
is unique and unambiguous as justified in \cite{BCST1}. Its properties are
illustrated in the plots in Fig.2b and Fig.3.

In Fig.3 the plots are for the gauge invariants $\chi$ and $\sigma.$ Recall
from Eqs.(\ref{chi},\ref{link1}) that $a_{E}^{2}\left(  \tau\right)
=\left\vert \chi\right\vert .$ Hence the crunch/bang occur when $\chi=0,$ and
it is seen from Fig.2a that the universe goes smoothly through these
singularities while undergoing transitions between the gravity and antigravity
regimes. It should be emphasized that this behavior is dictated by geometry
(anisotropy), it is model independent, unavoidable and unambiguous. Note also
that the duration of antigravity decreases with an increase in radiation or a
decrease in the kinetic energies of the scalar fields or anisotropy.

We have asked the question whether it is possible to avoid the antigravity
region? The answer is different depending on the presence or absence of
anisotropy. With non-zero anisotropy, no matter how small, it is not possible
to avoid the antigravity region, due to the unique solution illustrated in
Fig.2b and Fig.3. Since it is unlikely that anisotropy is identically zero
near the singularity, the conclusion is that, at the classical level,
antigravity cannot be avoided.

Nevertheless, we may still answer the question when anisotropy is identically
zero. Then the answer is yes, antigravity can be avoided, and still have a
geodesically complete universe in the Einstein frame, confined only to the
gravity patches in Fig.1. This is described by a non-generic subset of special
solutions that are obtained by restricting the parameter space or initial
conditions, and they describe zero-size and finite-size bounces, as shown in
Figs.(4-7) and Figs.(8,9) respectively. As seen in Figs.(6,7), the
trajectories of the zero-size bounces never reach into the antigravity sector
and they never cross the dotted diagonals except at the origin.

\begin{center}%
\begin{tabular}
[c]{ll}%
{\parbox[b]{3.3797in}{\begin{center}
\includegraphics[
height=2.1015in,
width=3.3797in
]%
{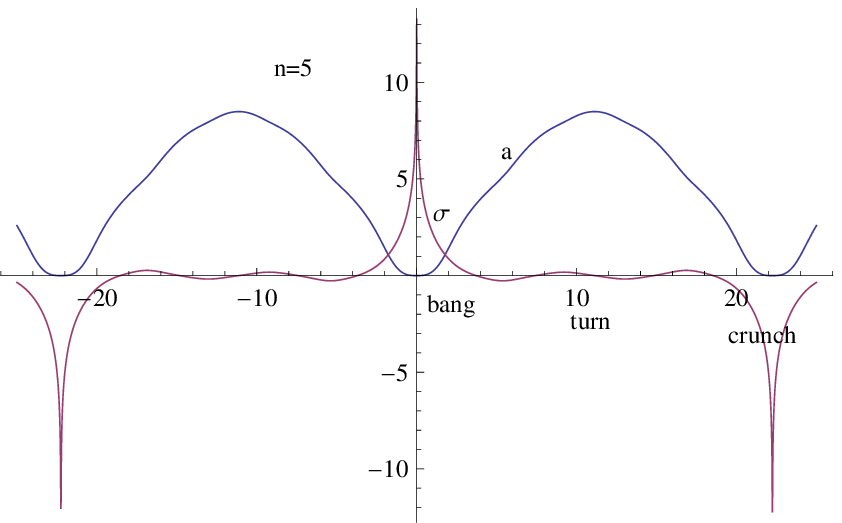}%
\\
Fig.4 - Zero-size bounces in a cyclic universe. $a_{E}\left(  \tau\right)  $
and $\sigma\left(  \tau\right)  $ for $b<0$ are depicted. In this example of
an analytic solution for the case of $V\left(  \sigma\right)  =b\cosh
^{4}\sigma+c\sinh^{4}\sigma,$ the fields $a_{E},\sigma$ are computed through
Eq.(\ref{link1}), so the behavior at crunch/bang times is consistent with
Fig.5. The number $n=5$ corresponds to the number of times $\sigma$ crosses
zero between a bang and a crunch, and is related to the qantization of periods
described in Fig.5. For $b>0$ (not shown) turnaround is at $a_{E}%
\rightarrow\infty.$%
\end{center}}}
&
{\parbox[b]{3.3797in}{\begin{center}
\includegraphics[
height=2.1015in,
width=3.3797in
]%
{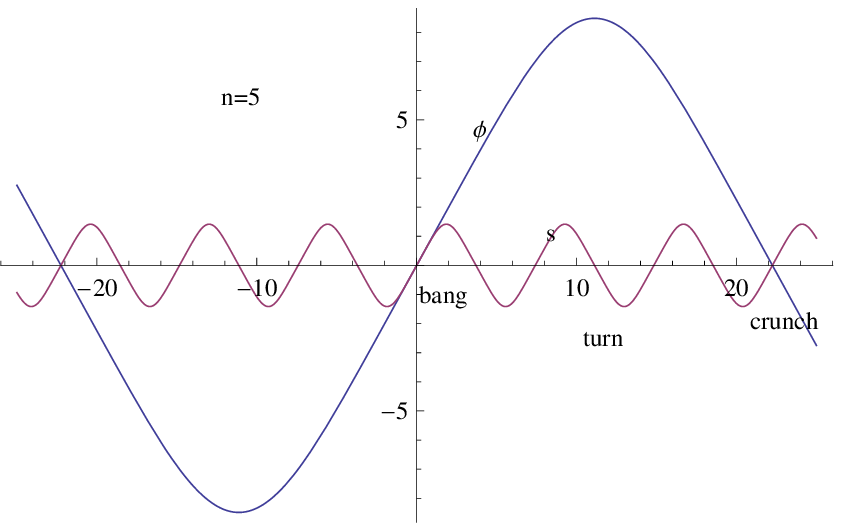}%
\\
Fig.5 - Zero-size bounces in a cyclic universe. $\phi_{\gamma}\left(
\tau\right)  $ and $s_{\gamma}\left(  \tau\right)  $ for $b<0$ are depicted.
In this example of an analytic solution for the case of potential energy
$\phi^{4}f\left(  s/\phi\right)  =b\phi^{4}+cs^{4},$ with $b<0$, the fields
$\phi_{\gamma}\left(  \tau\right)  $,$s_{\gamma}\left(  \tau\right)  $ have
synchronized initial values, $\phi_{\gamma}\left(  0\right)  =s_{\gamma
}\left(  0\right)  =0,$ and relatively quantized half periods (6 to 1 as seen
in the figure) so that $\phi_{\gamma},s_{\gamma}$ vanish simultaneously at all
cyclic crunch/bang times. For $b>0$ (not shown) turnaround is at $\phi
_{\gamma}\rightarrow\infty.$ The same solution is potted in Figs.6,7 in a
different way.
\end{center}}}
\end{tabular}

\ $\;\;\;\ \;$
\end{center}

The zero-size-bounce in Figs.(4-7) describes a cyclic universe that
\textit{contracts to zero size and then bounces back smoothly from zero size}.
They are exact analytic solutions of the cosmological equations for the case
of $\phi^{4}f\left(  s/\phi\right)  =b\phi^{4}+cs^{4},$ or equivalently
$V\left(  \sigma\right)  =b\cosh^{4}\sigma+c\sinh^{4}\sigma.$ The universe
expands up to either a finite size (when $b<0$) or infinite size (when $b>0$)
and then turns around to repeat the cycle. In these solutions, as described in
detail in \cite{BCT1}\cite{BCST2}, the behavior of $a_{E}^{2}\left(
\tau\right)  $ near the singularity at $\tau\sim0$ is smooth (if $\rho_{r}$ or
$k$ take generic values, then $a_{E}^{2}\left(  \tau\right)  \sim\tau
^{2}\rightarrow0$; if both $\rho_{r},k$ vanish or take some special values,
then $a_{E}^{2}\left(  \tau\right)  \sim\tau^{6}\rightarrow0).$ Also, the
behavior of the potential and kinetic energy terms for the scalar field
$\sigma\left(  \tau\right)  $ near the singularity are surprising. Namely,
contrary to the generic solution, the potential energy dominates over the
kinetic energy so that the equation of state $w\left(  \tau\right)  $ is
negative near the singularity. According to common lore, it was thought that
such zero-size-bounce solutions would not exist because they would violate the
null energy condition (NEC). However, this is not the case. The NEC is
satisfied because there is a singularity at zero size, and this is the
exception allowed according to the NEC theorems. To emphasize, the singularity
theorems are not violated, there is a singularity in the Einstein frame, but
what was missed is that it is possible to have a bounce of zero size, as
explicitly given analytically in our papers. We found all such solutions and
classified them in \cite{BCT1}\cite{BCST2}, thus providing a rich class of
examples of zero-size-bounce cyclic universes, characterized by arbitrary
values of the parameters $\rho_{r},k,b,c$ plus one additional quantized
parameter ($n$ in Figs.4-7).

To make the zero-size-bounce happen, a synchronization of initial conditions
and a quantization condition among the 6 available parameters must be imposed.
These properties are illustrated in Fig.5, where it is seen $s_{\gamma}\left(
0\right)  =\phi_{\gamma}\left(  0\right)  =0$ is imposed as an initial
condition, and the periods of oscillations of $\phi\left(  \tau\right)  $ and
$s\left(  \tau\right)  $ are quantized relative to each other. Hence such
solutions are characterized by 4 continuous and 1 quantized parameter, rather
than the 6 continuous parameters of the generic solutions. In that sense the
zero-size-bounce solutions are a set of measure zero. So, statistically, it
does not seem likely that the universe would choose such a solution over the
generic solution.%

\[%
\begin{array}
[c]{cc}%
{\parbox[b]{3.3797in}{\begin{center}
\includegraphics[
height=0.5898in,
width=3.3797in
]%
{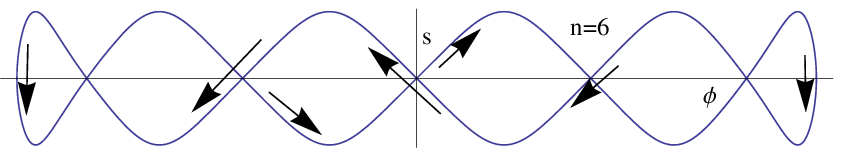}%
\\
Fig.6 - Geodesically complete trajectory only in the gravity sector $%
\phi^2-s^2\geq0$. For $b<0$ turaround is at finite $\phi_\gamma$. The number
of times ($n=6)$ the curve crosses the axis between a bang and a crunch is
determined by the quantization of the periods explained in Fig.5.
\end{center}}}
&
{\parbox[b]{3.3797in}{\begin{center}
\includegraphics[
height=0.3027in,
width=3.3797in
]%
{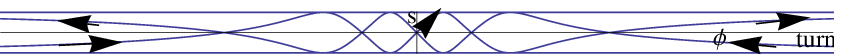}%
\\
Fig.7 - Geodesically complete trajectory only in the gravity sector, $%
\phi^2-s^2\geq0$. For $b>0.$ turnaround is at $\phi_\gamma=\infty$. The number
of times the curve crosses the axis between a bang and the next crunch is
determined by the relative quantization of the periods for $\phi_\gamma\left(
\tau\right)  $ and $s_\gamma\left(  \tau\right)  .$
\end{center}}}
\end{array}
\]

However, anisotropy plays a very important role in providing an attractor
mechanism such that, for typical initial conditions away from the singularity,
all trajectories are attracted to the origin of the $\left(  \phi,s\right)  $
plane and can cross the lightcone in Fig.1 only at the origin. In that sense
the type of solution depicted in Figs.(4-7) is not too far from being generic
once anisotropy is taken into account. In particular the behavior away from
the origin is a good approximation. Nevertheless, what goes on at the
singularity is quite different than what is depicted in Figs.(4-7). Namely,
antigravity cannot be avoided since the non-generic solutions of Figs.(4-7)
become strongly distorted by anisotropy, so that near the singularity they
approach the point $\phi_{\gamma}=s_{\gamma}=0$ \textit{tangentially} to the
dotted diagonal lines, with the same absolute rate $\left\vert \dot{\phi
}_{\gamma}\left(  0\right)  \right\vert =\left\vert \dot{s}_{\gamma}\left(
0\right)  \right\vert $ (which is different when anosotropy is absent), and
must dive into the antigravity region just as depicted in Fig.(2b,3), which is
different than Figs.6,7.

In the absence of anisotropy there are also non-generic finite size solutions.
The analytic solutions for these are given in \cite{BCT1}\cite{BCST2}, and an
example is depicted in Figs.(8.9). These are also repetitive as a function of
time, although not fully cyclic in the sense that the minimum size of the
universe (and other properties) could change from cycle to cycle.

\begin{center}%
\begin{tabular}
[c]{ll}%
{\parbox[b]{3.3797in}{\begin{center}
\includegraphics[
height=2.1015in,
width=3.3797in
]%
{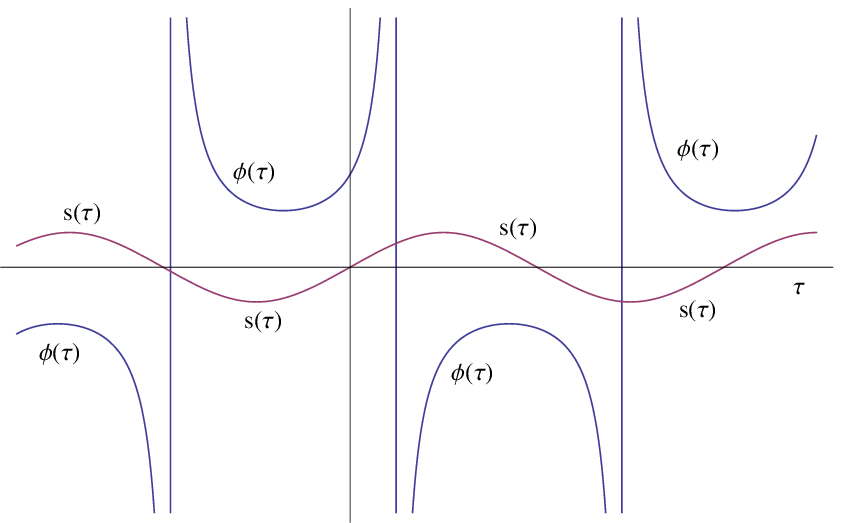}%
\\
Fig.8 - Finite-size-bounce, $\left\vert \phi_{\gamma}\right\vert >\left\vert
s_{\gamma}\right\vert $ for all $\tau$. It is possible only in a narrow range
of parameter space, including a condition on inital values in the range
$\phi_{\text{min}}^{2}>\left(  k/4b\right)  >s_{\text{max}}^{2}.$%
\end{center}}}
&
{\parbox[b]{3.3797in}{\begin{center}
\includegraphics[
height=2.1015in,
width=3.3797in
]%
{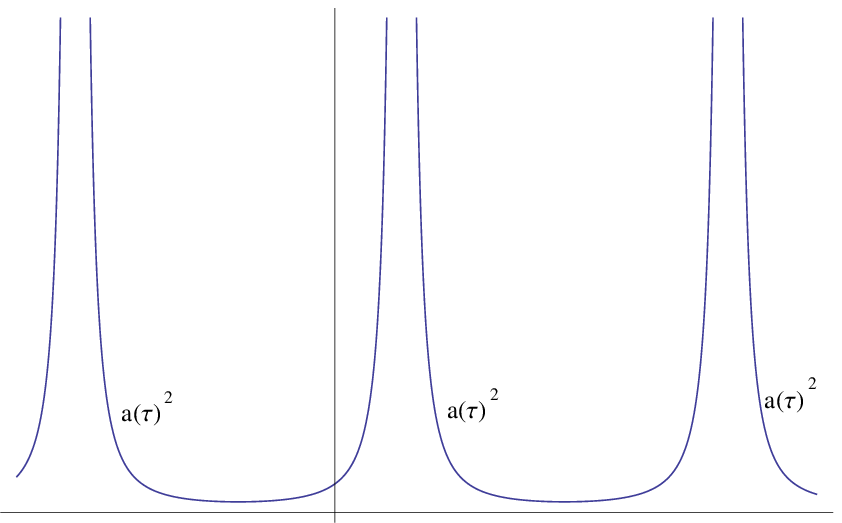}%
\\
Fig.9 - Finite-size-bounce. The scale factor, $a_{E}^{2}\left(  \tau\right)
=\left\vert \phi_{\gamma}^{2}-s_{\gamma}^{2}\right\vert ,$ never vanishes as
seen from Eq.(\ref{link1}) and Fig.8. The minumum size of $a_{E}^{2}\left(
\tau\right)  $ varies from cycle to cycle.
\end{center}}}
\end{tabular}

\end{center}

The finite-size-bounce describes a universe that contracts up to a minimum
non-zero size and then bounces back into an expansion phase up to infinite
size. As the universe turns around to repeat such cycles the minimum size is
not necessarily the same, as this depends on the parameters and initial
conditions. Such solutions occur when the parameters satisfy, $\rho_{r}%
<k^{2}/16b$ and $\phi_{\text{min}}^{2}\left(  \tau_{0}\right)
>k/4b>s_{\text{max}}^{2}\left(  \tau_{0}\right)  $; for details see
\cite{BCST2}. Note that there are still 6 parameters, so this is not a set of
measure zero, but it is a restricted small region of parameter space or
initial values.

In the presence of anisotropy these solutions are also distorted. However,
since $a_{E}\left(  \tau\right)  $ never vanished when anisotropy was
neglected, the effects of anisotropy may be less important. We have not fully
determined if, depending on initial conditions, it may be possible to avoid
the antigravity region for all values of the conformal time $\tau$. Even if
this is possible mathematically, since it corresponds to a very small region
in the space of parameters of the model and initial conditions, it seems it is
an unlikely physical scenario.

We are, therefore, faced with trying to understand physical phenomena in the
antigravity regime. Some relevant ideas are found in \cite{BCST1}\cite{BCST2}.
These include comments on negetive energy gravitons, an analogy to the
Klein-Paradox, and an outline on analytic continuations to circumvent the
singularity to connect data from before the crunch to after the bang, etc..
The difficult, but ultimately necessary, ingredient is the inclusion of the
full, possibly non-perturbative, quantum effects, that remain as a challenge.
However, progress and physical insight into the meaning and measurable effects
of antigravity are likely to develop well before understanding the full theory
of quantum gravity.

\section{Conclusions and Outlook}

\label{discussion}

This paper summarized our approach with analytic solutions of cosmological
equations to discuss geodesic completeness through the big bang singularity.
This analysis was done in the purely classical theory of gravity, but has some
immediate extensions to the quantum theory. In particular, in the context of
the path integral, our complete set of classical solutions provide a
semi-classical approximation to the quantum theory of gravity (like instantons
in QCD). Furthermore, in the context of string theory, our complete set of
solutions provide cosmological backgrounds consistent with perturbative
conformal invariance on the worldsheet, and hence a starting point for a
stringy investigation of quantum gravity in a cosmological setting.

We learned that the generic solution shows that the trajectory of the universe
goes smoothly through the crunch/bang singularities while traversing from
gravity to antigravity spacetime patches, and doing this repeatedly in a
periodic manner. The generic trajectory can cross the \textquotedblleft
lightcone\textquotedblright\ in field space, as shown in Fig.(2a), at any
place, if anisotropy vanishes. The crossing points on the \textquotedblleft
lightcone\textquotedblright\ depend on the parameters of the model and initial
conditions. When anisotropy is present, there is an attractor mechanism that
distorts the generic solution close to the singularity to the form shown in
Fig.2b and Fig.3. This is model independent, unique and unavoidable at the
classical level. Therefore, the phenomenon of antigravity should be considered
seriously in discussing cosmology.

It is possible to avoid antigravity and still have a geodesically complete
geometry, but only if anisotropy vanishes identically, which is an unlikely
scenario for the initial stages of a small universe, and only within a smaller
subset of initial conditions, which is of measure zero in the space of all
initial conditions. If these unjustifiable conditions are assumed, then the
zero-size bounces of the type in Figs.(4-7) and the finite size-bounces of the
type in Figs.(8,9) are the only geodesically complete solutions contained
totally within the traditional Einstein frame. One group of trajectories
passes through the center of the \textquotedblleft lightcone\textquotedblright
repeatedly, resulting in a cyclic universe as in Figs.(4-7). These solutions,
which do not violate the null energy condition, provide a set of examples that
bouncing at zero size is possible classically in cosmological scenarios with
or without spatial curvature.

We have shown that antigravity is very hard to avoid generically in the
classical theory. Anisotropy + radiation + kinetic energy of the scalar field
$\sigma,$ require the antigravity epoch. For close to a year we tried to find
models and mechanisms to avoid antigravity (i.e. when \textit{all} solutions
of a geodesically complete model are included). The failure to find such
mechanisms finally led us to take antigravity seriously. We also studied the
Wheeler-deWitt equation to take into account some quantum effects for the same
system that we analyzed classically. We could solve some cases exactly, others
semi-classically. We arrived substantially to the same conclusions that we
learned by studying the purely classical system.

The source of antigravity is the factor $\left(  \phi^{2}-s^{2}\right)
R\left(  g\right)  /12$ that can switch sign and becomes negative in the
antigravity regimes indicted in Fig.1. This factor is a consequence of the
Weyl symmetry as explained following Eq.(1). It cannot be replaced by the
absolute value $\left\vert \phi^{2}-s^{2}\right\vert $ to avoid the sign
change because this is not Weyl invariant. There are generalizations of this
factor consistent with Weyl symmetry as given in footnote (1) and
\cite{2tSugra}, but after taking into account additional features, such as
supersymmetry, there are no variations of this factor that would not switch
sign. Note from Eqs.(\ref{Egauge},\ref{link1}) that the signature of the
metric does \textit{not} change when when this factor switches sign. In
1T-physics Weyl symmetry is not required, so gravity can be formulated without
the antigravity sector, but as described following Fig.1, such a formulation
of gravity is necessarily geodesically incomplete, and therefore problematic.
By contrast, in 2T-physics Weyl symmetry is a consequence of just the presence
of the extra $\left(  1+1\right)  $ dimensions since Weyl symmetry amounts to
a reparametrization of the $\left(  1+1\right)  $ coordinates as part of the
general coordinate invariance in $4+2$ dimensions (note that Weyl symmetry is
not a symmetry of 2T-gravity \cite{2Tgravity}\cite{2TgravityGeometry} on top
of general coordinate invariance in $4+2$). So, Weyl symmetry may be
considered a feature and signature of 2T-physics.

It should be emphasized that our new results transcend the specific simple
models for which we found the complete set of analytic solutions. The
phenomena we have found should also be expected generically in supergravity
theories coupled to matter whose formulation include a similar factor that
multiplies $R\left(  g\right)  .$ In supergravity, that factor has the form
$\left(  1/16\pi G-K\left(  \varphi_{i},\varphi_{i}^{\ast}\right)  \right)
R\left(  g\right)  ,$ where the function $K\left(  \varphi_{i},\varphi
_{i}^{\ast}\right)  $ of the compex fields $\varphi_{i}$ is called the
K\"{a}hler potential. In the past, it was assumed that the overall factor in
front of $R\left(  g\right)  $ is positive, and investigations of supergravity
proceeded, by fiat, only in the positive regime $\left(  1/16\pi G-K\left(
\varphi_{i},\varphi_{i}^{\ast}\right)  \right)  >0$ \cite{weinberg}. For
example, a discussion of the field space in the positive sector for general
$\mathcal{N}$=2 supergravity can be found in \cite{deWit}. However, our
results suggest that generically the overall factor can and will change sign
dynamically, in every gauge, and therefore in the \textit{geodesically
complete supergravity}, antigravity sectors similar to our discussion in this
paper should be expected in typical supergravity theories (this is illustrated
with an example in \cite{BCST3}).

As emphasized above, in supergravity, the factor in front of $R\left(
g\right)  $ is not generally positive definite. To clarify this point in the
context of our formalism, we could chose a gauge in our simple theory to make
it look like supergravity. In a Weyl gauge that we call the supergravity
gauge, or $c$-gauge \cite{2Tgravity}\cite{BCST2}, in which $\phi(x)$ is set to
a constant $\phi_{0}$, our gauge fixed term $(\phi_{0}^{2}-s^{2})R(g)/12$
reduces precisely to the familiar form in supergravity, including a K\"{a}hler
potential. As we have argued, if $\left(  \phi^{2}-s^{2}\right)  $ changes
sign in one gauge, it must change sign in every gauge, because the sign is
gauge invariant under Weyl transformations. Hence, the similar factor in
supergravity is also expected to change sign. This can be made more evident in
a Weyl invariant reformulation of supergravity. This is not straightforward
because scalar fields in supergravity are complex, so a complex $\phi$ would
have an additional ghost that needs removal with a larger gauge symmetry
beyond the Weyl symmetry. I have shown in \cite{2tSugra} that the general Weyl
invariant approach can in fact be extended to supergravity where an additional
U$\left(  1\right)  $ gauge symmetry exists to remove the ghosts of a complex
$\phi$ field. Similarly, in supergravities with more supersymmetries $\left(
\mathcal{N}\geq2\right)  $ there are additional hidden gauge symmetries that
enlarge the hidden gauge group, which, together with the Weyl symmetry can
remove ghosts like $\phi$; this allows the inclusion of more fields like
$\phi$ that further generalize the factor analogous to $\left(  \phi^{2}%
-s^{2}\right)  R\left(  g\right)  /12,$ and thus permitting a richer variety
of useful gauge choices. So, the form $\left(  1/16\pi G-K\left(  \varphi
_{i},\varphi_{i}^{\ast}\right)  \right)  R\left(  g\right)  $ that occurs in
$\mathcal{N}=1$ supergravity is just the $c$-gauge fixed version of the
supersymmetric Weyl invariant supergravity formalism; this was derived again
from a $4+2$ dimensional 2T-supergravity \cite{2tSugra}. There is however one
additional constraint placed on $K\left(  \varphi_{i},\varphi_{i}^{\ast
}\right)  $ as a result of the underlying $4+2$ dimensional structure of
2T-supergravity: this constraint on scalar fields, as given in \cite{2tSugra},
which applies in all fundamental theories of physics, including the Higgs
field in the Standard model, may be taken as another feature and signature of
2T-physics which may show up in cosmology and/or accelerator physics.

Until better understood in the context of quantum gravity, or string theory,
our results should be considered to be a first pass for the types of new
physics questions they raise and the answers they provide. Much remains to be
understood, including quantum gravity and string theory effects, but it is
clear that previously unsuspected phenomena, including antigravity, come into
play classically close to the cosmological singularity. The technical tools to
study such issues in the context of a full quantum theory of gravity are yet
to be developed. This is an important challenge, since the results have
profound implications for both fundamental physics and our understanding of
the origin, evolution and future of the universe.

\bigskip

\begin{acknowledgments}
This research was partially supported by the U.S. Department of Energy under
grant number DE-FG03-84ER40168. I would like to thank my collaborators
Shih-Hung Chen, Paul Steinhardt and Neil Turok with whom this research was conducted.
\end{acknowledgments}

\end{document}